# Growth of Superconducting $MgB_2$ Thin Films

Kenji Ueda, Michio Naito
NTT Basic Research Laboratories, NTT Corporation,
3-1 Wakamiya, Morinosato, Atsugi, Kanagawa 243-0198, Japan

## 1. Introduction

The recent discovery of superconductivity at 39 K in $MgB_2$ [1] has generated great interest in both basic science and practical applications. $MgB_2$ has the highest superconducting transition temperature ($T_C$) among non-oxide materials and its $T_C$ is close to that of $La_{2-x}Ba_x$(or $Sr_x$)$CuO_4$, which were the first high-$T_C$ superconductors. The $T_C$ is slightly higher than the theoretical upper limit predicted for phonon-mediated superconductivity [2], which had been widely accepted until the discovery of superconducting cuprates. Therefore it is important to clarify the superconducting pairing mechanism of $MgB_2$, and many researchers have been working on this subject in variety of ways. The B and/or Mg isotope effect [3, 4] exhibited a substantial isotope exponent ($\alpha$ : ~ 0.30) for B and a small exponent ($\alpha$ : ~ 0.02) for Mg, indicating that B atom vibrations are involved significantly in the superconductivity of $MgB_2$. Photoemission spectroscopy [5], scanning tunneling spectroscopy [6], and NMR [7] also showed that $MgB_2$ is essentially a conventional s-wave BCS superconductor, although subtle problems remain related to the character (single-gap or multiple-gap) and magnitude of the superconducting gap [8, 9].

With a view to its application, $MgB_2$ has been prepared in various forms such as bulks, wires and films. Of these, superconducting films are particularly important for electronics applications such as Josephson junctions and superconducting quantum interference devices (SQUID). $MgB_2$ is a promising material for preparing junctions because it has a simpler crystal structure, fewer material complexities, and a longer coherence length (ξ: ~ 5 nm) than cuprates in spite of its lower $T_C$ [10]. In this article, we review the efforts (including our work) made to grow films in the year that has passed since the discovery of $MgB_2$. We also discuss the prospects for Josephson junction fabrication using $MgB_2$.

## 2. Thin films

There are two complicated problems related to the preparation of superconducting $MgB_2$ films: the high sensitivity of Mg to oxidation and the large difference between the vapor pressures of Mg and B. We can avoid the former problem by depositing Mg and B from pure metal sources in an ultra high vacuum chamber. The latter problem is more serious, and there are two methods for solving it. One is to prepare films under a high Mg vapor pressure in a confined container at high temperatures, and the other is to prepare films at low temperatures. The first method is employed in "two-step" synthesis, in which the first step is the deposition of amorphous B (or Mg-B composite) precursors, and the second step involves annealing at elevated temperatures with Mg vapor usually in an evacuated Nb, Ta or quartz tube. The second method is employed in "as-grown" synthesis. Each method has its merits and demerits. Two-step synthesis produces good crystalline films with superconducting properties comparable to those of high-quality sintered specimens although it cannot be used to fabricate Josephson junctions or multilayers. By contrast, as-grown synthesis can produce only poor crystalline films that have slightly lower $T_C$ (typically 35 K) than the bulk value, but this approach makes multilayer deposition feasible. Two-step synthesis is described in 2.1 and as-grown synthesis in 2.2. The substrate effect is summarized in 2.3. Two interesting topics are touched on in 2.4.

### 2. 1. Two-step synthesis

Most of the initial attempts to fabricate $MgB_2$ films employed two-step synthesis using the pulsed laser deposition (PLD) technique to prepare amorphous precursors. PLD was used not because it is especially advantageous, but because it found popularity in relation to the easy preparation of complex oxide films including high- $T_C$ cuprates. In principle, however, two-step growth can be achieved by any thin-film preparation technique including sputtering, and E-beam (or thermal) evaporation.

The two-step synthesis process is as follows [11-16]. First, amorphous B (or Mg-B composite) precursors are deposited on substrates, usually at ambient temperature, by using PLD, sputtering or E-beam (or thermal) evaporation. The film thickness is typically 400-500 nm. The film thickness is important to a certain extent because there is interdiffusion between films and some substrates above 600°C [17]. Then, the precursors are sealed in evacuated (or sometimes Ar-containing) Nb or Ta tubes with Mg metal pieces as shown in Fig. 1. In some cases, the precursors are wrapped in Nb or Ta envelopes with Mg metal pieces and then this is all encapsulated in an evacuated quartz tube. The amount of Mg has to provide an Mg vapor pressure sufficient to form $MgB_2$ at the annealing temperature. Nb or Ta tubes generally give better results than bare quartz tubes since Mg vapor severely attacks quartz tubes at elevated temperatures. Finally the tube is heated at 600 - 900°C for 10 to 60 min. in an outside furnace (*ex-situ* annealing). The annealing profile differs for each group. Several groups claim that rapid heating to the annealing temperature and quenching to room temperature may be important, while other groups do not. Table I summarizes the film preparation recipes and physical properties of the resultant films reported by several groups that have used two-step synthesis to produce high-quality films. Early films prepared by two-step synthesis often contained a small quantity of MgO impurities, which can be significantly reduced by taking great care to avoid air exposure at every step of the process. It has been reported that similar caution is needed in two-step synthesis of high-quality Tl- or Hg-based superconducting cuprate films [18].

Of the groups employing two-step synthesis, a group at Pohang University of Science and Technology has obtained the best results. Their films are highly crystallized as suggested by their XRD peak intensity (Fig. 2) [11]. The films are [101] oriented on $SrTiO_3$ (100) substrates and *c*-axis oriented on sapphire ($Al_2O_3$) -R. The superconducting transition temperature ($T_C$) is 39 K at zero resistance, which is the same as the bulk $T_C$ (Fig. 3). The residual resistivity ratio (RRR) is around 3. Their $T_C$ and RRR are the best of all films grown to date. The RRR is only slightly smaller than the value of 5 ~ 6 reported for single crystals [19]. The RRR of most of the films prepared by other groups is only slightly above 1. As regards the critical current densities ($J_C$), the Pohang group achieved values of ~ 40 MA/cm$^2$ at 5 K and 0 T, and ~0.1 MA/cm$^2$ at 15 K and 5 T [20]. Similar values were reported by Eom *et al*. and Moon *et al* [12, 14]. Furthermore, very low surface resistance ($R_S$) values (<100 μΩ at 4.2K and 270 μΩ at 15 K and 17.9 GHz) were obtained for the films prepared by the Pohang group [21]. These $J_C$ and $R_S$ values suggest that $MgB_2$ films are promising for practical applications.

Some groups have succeeded in preparing superconducting $MgB_2$ films by annealing precursor films "*in-situ*" in a growth chamber [15, 16, 22-28]. In this process, the most serious problem is the limited Mg vapor pressure inside a vacuum chamber. So the annealing recipe is significantly different from that used in the "*ex-situ*" annealing process: the annealing temperature is lower, typically ~ 600°C, and the annealing time is shorter, typically a few to 20 min. Even with such quick low-temperature annealing, films can often lose Mg, resulting in poor superconducting properties. To compensate for the Mg loss during *in-situ* annealing, precursor films are usually very rich in Mg. In some cases [25], an extra Mg cap layer is deposited on the top of Mg-enriched precursor films. In other cases [22, 23, 26], films are exposed during annealing to Mg plasma that is generated by ablating Mg metal by PLD (Mg plasma annealing). One important conclusion reached by most of the groups that used PLD for precursor deposition is that oxygen contamination (namely the amount of MgO) in the precursors must be minimized in order to produce superconducting films. It has been claimed that precursor films should be prepared in a blue laser plume (characteristic color of metal Mg) rather than a green plume (color of MgO) [22, 23, 26]. The *in-situ* plasma annealing also has to be performed in blue plasma. Table II summarizes the film preparation recipes and physical properties of the resultant films obtained by two-step synthesis employing "*in-situ*" annealing. Most of the films prepared by this process have an RRR of around 1 and a $T_C$ of around 25 K, that is significantly lower than the bulk $T_C$ or the $T_C$ of "*ex-situ*" annealed films. Zeng *et al*., however, got fair quality of films with a $T_C$ of around 34 K and an RRR of around 1.4 (see Fig. 4) [23].

2. 2. As-grown films

It is now one year since the discovery of superconducting $MgB_2$, and, to the best of our knowledge, only four groups including ours have reported the preparation of as-grown superconducting $MgB_2$ films [29-32]. This contrasts with the case of two-step synthesis, for which a fair number of reports have already appeared. Table III summarizes the film preparation recipes and physical properties of films grown by as-grown synthesis. Two groups ([29] and [32]) prepared films by coevaporation in an UHV chamber, one group by sputtering, and another by PLD. In this subsection, we first introduce our results, and then move on to the results of other groups.

We grew $MgB_2$ films in an MBE system that has been used for the last decade to grow superconducting cuprate films. Figure 5 shows a schematic diagram of our custom-designed MBE system [33, 34]. The base pressure of our system is ~ $1\text{-}2\times10^{-9}$ Torr, and the pressure during the growth is ~ $5\times10^{-8}$ Torr. Pure Mg and B metals were deposited by multiple electron beam evaporators. The evaporation beam flux of each element was controlled by electron impact emission spectrometry (EIES) via feedback loops to the electron guns. We varied the flux ratio of Mg to B from 1.3 to 10 times the nominal flux ratio to avoid the Mg loss. The growth rate was 0.2 -0.3 nm/s, and the film thickness was typically 100 nm. The growth temperature ($T_S$) ranged from 83°C to 650°C. The substrates used are $SrTiO_3$ (100), sapphire -R, sapphire -C, and H-terminated Si (111).

As mentioned above, the main problem when preparing as-grown $MgB_2$ films is the high volatility of Mg. Figure 6 shows the molar ratio of Mg to $B_2$ for films grown at various temperatures obtained by inductively coupled plasma spectroscopy (ICP) analysis. Films grown above 350°C were significantly deficient in Mg even with a 10 times higher Mg rate. This seems to be the growth temperature limit in terms of preparing as-grown superconducting $MgB_2$ films by this method. Films grown below 300°C may contain excess Mg especially if grown with a 10 times higher Mg rate. This excess Mg may affect the resistivity of the film, however, it should not affect $T_C$.

We studied the crystal structures of the $MgB_2$ thin films using $2\theta$-$\theta$ X-ray diffraction (XRD). We observed no peaks (except substrate peaks) for films formed on $SrTiO_3$ (100) and sapphire -R substrates, which have square- or rectangular surface structures. These films had halo (or sometimes ring) RHEED patterns. By contrast, for films formed on Si (111) and sapphire -C substrates, which have hexagonal surface structures, we were just able to observe tiny peaks. These peaks can be attributed to $MgB_2$ (00*l*) peaks (Fig. 7). The RHEED showed spot patterns indicating a single crystalline nature. These results show that films grown on sapphire -R and $SrTiO_3$ (100) are amorphous or polycrystalline, while films on Si (111) and sapphire -C have a *c*-axis preferred orientation although their crystallinity is very poor. A large difference can be seen between the XRD peak intensities of annealed and as-grown films (compare Figs. 2 and 7)

The films formed at $T_S$ of 150 to 320°C showed superconductivity in spite of their poor crystallinity. Figure 8 shows the resistivity versus temperature curves ($\rho$-$T$ curves) for $MgB_2$ films on Si (111) and sapphire -R as a function of the growth temperature ($T_S$). Films formed on Si (111) above 150°C showed superconductivity. The transition temperatures ($T_C$ onset -$T_C$ zero) were 12.2 - 5.2, 26.3 - 25.2, and 33.0 - 32.5 K for $T_S$ of 150, 216, and 283°C, respectively (Fig. 8(a)). The $T_C$ increased as the growth temperature increased. Films grown above 320°C and below 83°C were insulating or semiconducting, and did not show superconductivity. In case of films on sapphire -R, films formed at $T_S$ of 216 to 320°C showed superconductivity (Fig. 8(b)). It is surprising that $MgB_2$ films with poor crystallinity grown at $T_S$ of as low as 150°C show superconductivity although $T_C$ is depressed. This indicates that long range crystal ordering is unnecessary in order to achieve superconductivity in $MgB_2$. This is in contrast with the case of high-$T_C$ cuprates.

With regard to the substrate effect, Fig. 9 (a) compares the superconducting transition of the films on four substrates. The large scatter seen in the resistivity is due to residual Mg, the amount of which may differ for each film. The $T_C$ of films on Si (111) and sapphire -C is slightly higher than that on sapphire -R and $SrTiO_3$ (100). This trend holds over the whole growth temperature range of 150 – 320°C, as seen in Fig. 9(b), which shows the growth temperature dependence of $T_C$ on various substrates. The surfaces of Si (111) and sapphire -C are hexagonal, while those of sapphire -R and $SrTiO_3$ (100) are square or rectangular. $MgB_2$ has a hexagonal crystal structure with hexagonal Mg and B planes stacked alternately along the c axis [1]. Therefore, Si (111) and sapphire -C provide a better match with $MgB_2$ than sapphire -R and $SrTiO_3$ (100). Our results indicate that the use of hexagonal substrates leads to a slight improvement in the quality of the $MgB_2$ film.

The critical current density of one film on sapphire -R has been estimated by transport measurements. The value was $4.0\times10^5$ A/cm$^2$ at 4.2 K and 0 T. The $J_C$ of this film is comparable to that of bulk samples [10], but much smaller than that of films prepared by two-step synthesis (see Table I).

Jo *et al*. of the Stanford group also prepared $MgB_2$ films in an MBE chamber [32]. They obtained similar results to ours. They tried to find a convenient "window" for $MgB_2$ + Mg-gas based on the Mg-B binary phase diagram calculated by Liu *et al* [35], but they failed. Their results indicate that too much Mg flux results in $MgB_2$ + Mg-solid, and too little results in $MgB_2$ with Mg-deficit phases ($MgB_4$ etc.). Their optimized growth temperature is identical to ours, $T_S$ of 295-300°C. Although their highest $T_C$ (end) ~ 34 K is similar to ours, the crystallinity of the films appears to be better than ours. The XRD patterns show definite $MgB_2$ peaks. Furthermore, the transmission electron microscopy image shown in Fig. 10 are very similar to those obtained for films grown by two-step synthesis [12]. This figure indicates a microstructure consisting of 40 nm hexagonal grains with some texture toward in-plane alignment.

Saito *et al*. employed carrousel-type sputtering as shown in Fig. 11 [30]. The Mg and B metal targets were placed on two adjacent cathodes, and each sputtering power was controlled independently. The carrousel with the substrate holder rotated at 50 rpm during the deposition. Films prepared at substrate temperatures between 300°C and 400°C showed superconductivity. The best $T_C$ and RRR of their films were ~ 28 K (with a transition width of ~ 1 K) and 1.1, respectively.

Grassano *et al*. [31] prepared films by PLD with an Mg enriched target (Mg:B = 1:1) at $T_S$ ~ 450°C. The crucial factor with regard to obtaining superconducting films is that the growth should be undertaken in a blue laser plume (color of Mg metal plasma) achieved only in a narrow range (~ $2\times10^{-2}$ mbar) of Ar buffer gas pressure. The same statement has been reached in two-step synthesis with "*in-situ*" annealing as mentioned above [22, 23, 25]. The films showed a $T_C$ (onset) of ~ 25 K and $T_C$ (zero) of ~ 22.5 K.

All the above results indicate that the following two points are important in terms of obtaining as-grown superconducting $MgB_2$ films.

(1) The growth temperature must be kept lower than ~ 300-350°C to avoid significant Mg loss.

(2) Oxygen must be excluded during the growth because it seems to prevent $MgB_2$ to form at ~ 300-350°C. An ultra high vacuum seems to be preferable for the growth of as-grown $MgB_2$ films.

None of the as-grown films reported to date are single-crystalline or epitaxial. If we wish to form epitaxial films, certain ingredients must be added during the growth to promote migration at $T$s of ~ 300-350°C.

Finally we make a short comparison between as-grown synthesis and two-step synthesis with "*ex-situ*' or "*in-situ*' annealing. In Fig. 12, the $T_C$ of the film is plotted against the growth temperature ($T_S$) for each type of synthesis. The highest $T_C$ (as high as bulk $T_C$) is obtained by two-step synthesis with "*ex-situ*" annealing although $T_S$ is also the highest. A low $T_C$ of around 25 K (except for one results reported by Zeng *et al*.) is obtained by two-step synthesis with "*in-situ*" annealing, in which $T_S$ is intermediate. A $T_C$ of ~35 K, slightly lower than bulk $T_C$, is obtained by as-grown synthesis, in which $T_S$ is the lowest. This comparison indicates that the proper choice will be two-step synthesis with "*ex-situ*" annealing for high-quality films and as-grown synthesis for multilayer films. Two-step synthesis with "*in-situ*" annealing seems inappropriate for any purpose.

## 2. 3. Effects of the substrate

The choice of substrate is important in terms of achieving less interdiffusion and better lattice matching. The substrates generally used for $MgB_2$ film preparation are sapphire ($Al_2O_3$) -R, -C, Si (100), (111), $SrTiO_3$ (100), (111), MgO (100), and SiC (0001). $SrTiO_3$ (100), MgO (100), and sapphire -R substrates are most frequently used for the growth of $MgB_2$ films because they have been widely used for the growth of high- $T_C$ cuprates. Some researchers use Si, sapphire -C, and SiC.

As regards interdiffusion, He *et al*. examined the reactivity between $MgB_2$ and common substrate materials ($ZrO_2$, YSZ, MgO, $Al_2O_3$, $SiO_2$, $SrTiO_3$, TiN, TaN, AlN, Si and SiC) [17]. In their experiments, each of these substrate materials in fine powder form was mixed with Mg metal flakes and amorphous B powder, and reacted at elevated temperatures (600, 700 and 800°C). Elemental Mg and B were employed in these reactions, rather than preformed $MgB_2$, to provide a better model of the film fabrication process. Furthermore very fine powder was used to enhance reactivity even at temperatures as low as ~ 600°C, as often used in thin film preparation. The results are summarized in Table IV. Surprisingly, $MgB_2$ has been found to be rather inert to many substrate materials. Even at 800°C, no reaction occurs with $ZrO_2$, MgO, or nitrides (TiN, TaN, AlN). The exceptions are $SiO_2$ and Si, where there is a severe reaction at 600°C, and $Al_2O_3$, where a reaction is observed at 700°C. At 800°C, $MgB_2$ is also reactive with $SrTiO_3$ and SiC. These results are helpful in terms of selecting appropriate substrates for thin film device applications.

Next, we mention the lattice matching of $MgB_2$ and substrates. Although no group has yet succeeded in growing single-crystalline epitaxial films of $MgB_2$, these considerations will be helpful in the future. Table V summarizes the crystal structure and lattice constants of $MgB_2$ and several well-known substrates. Figure 13 also shows the lattice constants of various surfaces of $MgB_2$. SiC is the best when considering only simple lattice matching. When we take higher order lattice matching into account, Si may also be a good choice. According to the idea proposed by Zur and McGill [36], the criteria for possible heteroepitaxy are as follows: (1) a two-dimensional (2D) superlattice cell area of less than 60 $nm^2$ and (2) a lattice mismatch at the interface between 2D superlattice cells of less than 1%. This idea has been experimentally confirmed in CdTe (111) on $Al_2O_3$ (111), HfN (111) on Si (111), HfN (100) on Si (100), etc. [37, 38]. For $MgB_2$, the combination of 5×5 triangular unit cells of $MgB_2$ and 4×4 unit cells of Si (111) satisfies this criterion with a mismatch less than 1% as shown in Fig. 14.

### 2.4. Other interesting topics

In this last subsection, we touch on two interesting topics related to $MgB_2$ films. Hur *et al*. prepared $MgB_2$ films on boron single crystals by two-step synthesis, and observed $T_C$ enhancement (~ 41.7 K) [39]. As they pointed out, a possible explanation for this enhancement is tensile epitaxial strain between the $MgB_2$ films and the boron substrates although this has yet to be confirmed. Epitaxial strain at the interface is one method of enhancing the $T_C$ of thin films as demonstrated for high- $T_C$ cuprates [40].

Li *et al*. have succeeded in preparing $MgB_2$ films on stainless steel by two-step synthesis [41]. They used special precursors, a suspension of magnesium and amorphous boron mixed together by stirring them in acetone. Their method has some similarity to the sol-gel process used for high-$T_C$ cuprates [42]. The suspension is deposited on a stainless steel substrate several times. The acetone is evaporated each time until the desired thickness is reached. The resulting powders are pressed and capped with another piece of stainless steel. Then the samples are sintered at 660 - 800°C. They obtained $T_C$ ~ 37.5 K and $J_C$ ~ $8\times10^4$ at 5 K and 1 T. Unfortunately the $MgB_2$ films do not adhere well to the steel substrate probably due to large difference in their thermal expansion coefficients. Their method may provide a synthetic route for preparing $MgB_2$ coated conductors.

## 3. Junctions

Superconducting junctions are important in terms of both basic and application research. $MgB_2$ may be suitable for preparing junctions because it has less anisotropy, fewer material complexities, and a longer coherence length (ξ= ~ 5 nm) than cuprates in which reliable and reproducible processes for fabricating good Josephson junctions have not yet been established in the 15 years since their discovery [10, 43]. Several groups have made various types of superconducting junctions using $MgB_2$, including point contacts, break junctions, and nanobridges. [8, 44-51]. Table VI summarizes these results.

Zhang *et al*. employed a point contact method using two pieces of $MgB_2$, and obtained either SIS or SNS junctions by adjusting the contact pressure [47]. With loose contact, they obtained SIS characteristics that show fairly good quasiparticle spectra (Fig. 15). The spectra are in good agreement with the standard s-wave BCS curve, and yield an energy gap of 2.02±0.08 meV. However, the resultant $2\Delta/k_BT_C$ value of 1.20 is significantly lower than the predicted BCS value of 3.52, indicating a possible reduction in the superconducting gap at the surface. The value of the superconducting energy gap obtained for $MgB_2$ by using various measurement techniques, such as scanning tunneling spectroscopy and high-resolution photoemission spectroscopy, ranges from 0.9 to 7.0 meV, and there is currently no consensus on its value. Some groups have suggested two energy gaps (multi-gap feature) [8, 9, 44] although other groups have claimed a single gap [5, 6, 47 - 50].

Zhang *et al*. also obtained SNS characteristics by tight contact as shown in Fig. 16 [47]. The experimental data are in good agreement with the predictions provided by the resistively-shunted junction (RSJ) model. A DC SQUID made from two SNS junctions yielded magnetic flux noise and field noise as low as $4\mu\Phi_0$ $Hz^{-1/2}$ and 35 fT $Hz^{-1/2}$ at 19 K, where $\Phi_0$ is the flux quantum. The low-frequency noise is 2-3 orders of magnitude lower than that of the YBCO SQUID early in its development, indicating that $MgB_2$ has an excellent potential for providing a SQUID that operates around 20 - 30 K, which is easily reached by current commercial cryocoolers.

There have also been some reports of Josephson junctions and SQUIDs fabricated on $MgB_2$ thin films [44, 46, 47, 51]. Brinkman *et al*. fabricated a SQUID using nanobridges (30×5 μm) patterned by focused-ion-beam (FIB) milling on an $MgB_2$ film [46]. The device operated below 22 K with 30 μV modulation at 10 K. Carapella *et al*. fabricated sandwich-type Nb/ $Al_2O_3$/Al/ $MgB_2$ thin-film junctions, and confirmed the dc and ac Josephson effect [44]. All the above results, although not yet ideal, seem to indicate that there may be fewer problems involved in fabricating superconducting junctions with $MgB_2$ than with high-$T_C$ cuprates.

## 4. Summary

In this short review, we undertook a broad survey of the efforts made in the year since superconducting $MgB_2$ films was discovered. We also present the prospects for Josephson junction fabrication using $MgB_2$. $MgB_2$ seems to have an excellent potential for electronics applications because of its simple crystal structure, small anisotropy, and long coherence length. In terms of device applications, the preparation of high-quality, epitaxial, and as-grown films is highly desired but not yet achieved. We hope this review will help the efforts being made in this direction. Finally we apologize that we were unable to include many of important results in this article for the limitation of the space.

Acknowledgments

The authors thank Dr. S. Karimoto and Dr, H. Yamamoto for fruitful discussions, and Dr. H. Takayanagi and Dr. S. Ishihara for their support and encouragement throughout the course of our research.

Figure captions

Fig. 1: Schematic diagram of two-step synthesis employing ex-situ annealing.

Fig. 2: XRD patterns for $MgB_2$ thin films grown on (A) (100) $SrTiO_3$ (STO) and (B) $Al_2O_3$ (AO) substrates (from ref. [11]).

Fig. 3: Temperature dependence of resistivity of $MgB_2$ thin films for H=0 and 5 T. The lower inset shows a magnified view near the $T_C$. The upper inset is a schematic diagram of the Hall pattern (from ref. [11])

Fig. 4: (a) Resistivity vs. temperature curve for a 400-nm-thick $MgB_2$ film. (b) The ac susceptibility of the same film (from ref. [23]).

Fig. 5: Schematic diagram of our MBE system.

Fig. 6: The molar ratio of Mg (compared with that of $B_2$) in the films against the growth temperature. The ideal ratio is one (shown by bold line).

Fig. 7: X-ray $2\theta$-$\theta$ diffraction patterns of films formed on (a) Si (111) or (b) sapphire -C as a function of the growth temperature (83-283°C). Peaks corresponding to $MgB_2$ are labeled. Peaks labeled "sub." correspond to substrate peaks or false reflections and are also observed in the XRD patterns of bare substrates.

Fig. 8: Resistivity versus temperature (ρ-T) curves of $MgB_2$ films formed on (a) Si (111) and (b) sapphire -R as a function of growth temperature (83-283°C).

Fig.9: (a) ρ-T curves of films formed at 283°C on various substrates (Si (111), sapphire -R, sapphire -C and $SrTiO_3$ (001) (STO)). (b) The $T_C^{onset}$ of the film on Si (111) (open circle), sapphire -C (closed diamonds), sapphire -R (closed triangles) and $SrTiO_3$ (001) (open squares) as a function of the growth temperature.

Fig. 10: Plane-view TEM and diffraction pattern of $MgB_2$ (from ref. [32]).

Fig. 11: Schematic diagrams of the carrousel-sputtering method.

Fig. 12: Film's $T_C$ as a function of growth temperature for each synthesis.

Fig. 13: The lattice constants of various surfaces of $MgB_2$.

Fig. 14: Epitaxial relation between $MgB_2$ (001) (area: 5x5) and Si (111) (area: 4x4).

Fig. 15: Current (crosses) and conductance *dI/dV* (diamonds) vs. voltage for $MgB_2$ tunnel junctions with fit to the theory shown as solid and dotted curves. (a) Temperature 8.9 K and (b) temperature 16.4 K. The inset in (a) is $\Delta(T)$ vs. temperature curve (from ref. [47]).

Fig. 16: Current vs. voltage at 5 K for SNS junction (from ref. [47]).

Table I: Recipes for film preparation and the physical properties of films prepared by two-step synthesis employing "*ex-situ*" annealing.

| Ref. | Authors | Method | Pre-cursors | Sub-strates | Thick-ness (μm) | Temp. (°C) | Time | Anneal tube | $T_C$(K) | RRR | $J_C$(A/cm$^2$) | XRD orientation |
|---|---|---|---|---|---|---|---|---|---|---|---|---|
| 11 | Kang | PLD | B-film | $SrTiO_3$(100) | 0.4 | 900 | 10-30-min. | Ta | 39-37.6 | 2.5 | $6*10^6$ (5K, 0T) | (101) |
| | | | | $Al_2O_3$-R | | | | | | | | (001) |
| 12 | Eom | PLD | Mg+-B-film | $SrTiO_3$(111) | | 850 | 15min. | Nb | 36-34 | ~1 | $3*10^6$ (4.2K, 1T) | (001) |
| | | | | | | 750 | 30min | Ta in Nb | 34-30 | | | |
| | | | | | | 750 | 30min | Quartz only | 34-29 | | | |
| 13 | Paranthaman | E-beam | B-film | $Al_2O_3$-R | 0.5-0.6 | 890 | 10-20-min | Ta | 39-38.6 | | $2*10^6$ (20K, 0T) | (001) |
| 14 | Moon | E-beam | B-film | $Al_2O_3$-C MgO | 0.25-0.3 | 700--950 | 30min | Ta, Ti | 39.2-38.9 | 1.5 | $1.1*10^7$ (15K,0T) | Poly |
| 15 | Zhai | E-beam | B-film | $Al_2O_3$-R | | 900 | 1hr | Ta | 39-38.8 | | | |
| | | PLD | Mg+-B-film | | | | | | 28.6-25.2 | | | |
| 16 | Plecenik | Thermal | B-film | $Al_2O_3$ random | 0.15 | 800 | 30min | Nb | 39-37 | ~1 | | |
| | | | Mg+-B-film | | 0.1-0.2 | | | | 33.3-32 | | | |

Table II: Recipes for film preparation and the physical properties of films prepared by two-step synthesis employing "*in-situ*" annealing.

| Ref. | Authors | Method | Precursors | Substrates | Growth temp. (°C) | Time | $T_C$(K) | RRR | $J_C$(A/cm$^2$) | Others |
|---|---|---|---|---|---|---|---|---|---|---|
| 23 | Zeng *et al.* | PLD | Mg+$MgB_2$: 250-300deg | $Al_2O_3$-C | 630 | 10min | 35.5-34 | 1.4 | 1.3*$10^6$ (7.5K, 0T) | Mg plasma |
| 15, 24 | Zhai *et al.* | PLD | Mg+$MgB_2$:RT | Si(100) | 630 | 20min | 25-24 | <1 | | |
| 25 | Christen *et al.* | PLD | Mg-rich $MgB_2$+Mg cap | $Al_2O_3$-R | 600 | 20min | 26.5-2-2.5 | <1 | | Special targets Mg plasma |
| 22, 26 | Brinkman *et al.*, Blank *et al.* | PLD | Mg-rich $MgB_2$+Mgcap | Si(100) | 600 | few min | 20-16 | ~1 | | Mg plasma (Same group as Brinkman) |
| | | | | $SrTiO_3$(100) | | | 23-22.5 | 18 | | |
| | | | | SiC(0001) | | | 25-21.8 | ~1 | | |
| | | | | MgO(100) | | | 26-23.5 | ~1 | | |
| 27 | Shinde *et al.* | PLD | Mg-B multilayers | $SrTiO_3$(111) (100) | 900 | 30min | 22 | ~1 | | |
| 28 | Ermolov *et al.* | sputteri-ng | Mg-$MgB_2$ composit | $Al_2O_3$-R | 600 | few sec | 24 | ~1 | 10*$10^6$ (12K, 60Oe) | Smaller (32mm) targets |
| 16 | Plecenik *et al.* | thermal evapor-ation | Mg, B co-deposition | $Al_2O_3$ random | 900 | 30sec | 26-16 | ~1 | | |
| | | | | Si(100) | | | 27-17 | | | |

Table III: Recipes for film preparation and the physical properties of resultant films prepared by "as-grown" synthesis.

| Ref.. | Authors | Method | Source | Substrates | Growth temp.($^{o}C$) | $T_C$(K) | RRR | $J_C$($A/cm^2$) | Others |
|---|---|---|---|---|---|---|---|---|---|
| 29 | Ueda *et al.* | MBE | Mg and B metal | $Al_2O_3$-R, -C $SrTiO_3$, Si | 320 | 36-35.5 | ~2 | $4.0*10^5$ (4.2K, 0T) | Co-depo-sition |
| 30 | Saito *et al.* | sputter-ing | Mg, B two target | $Al_2O_3$-R | 252 | 27.8-27.3 | 1.1 | | Carrousel sputtering |
| 31 | Grassano *et al.* | PLD | Mg+B pressed pellet | | 450 | 25-22.5 | ~1 | | Mg plasma |
| 32 | Jo *et al.* | MBE | Mg and B metal | $Al_2O_3$-C | 295-300 | 34.5-34 | ~2 | $4.0*10^6$ (5K, 0T) | Co-depo-sition |

Table IV: Reactivity of $MgB_2$ with various electronic materials (from ref. [17]).

| Electronic material | 600 °C anneal | 800 °C anneal |
|---|---|---|
| $ZrO_2$ | No reaction | No reaction |
| YSZ[a] | No reaction | $MgB_2$, small amount of MgO |
| MgO | No reaction | No reaction |
| $Al_2O_3$ | No reaction | $MgB_2$ with altered cell size, MgO, unknown |
| $SiO_2$ | $MgB_2$, MgO, Si | $MgB_2$, $MgB_4$ MgO, $Mg_2Si$, Si |
| $SrTiO_3$ | No reaction | $MgB_2$, $SrTiO_3$, MgO $SrB_6$, $Ti_2B$ |
| Si | $MgB_2$, $Mg_2Si$ | $MgB_2$, $Mg_2Si$, $MgB_4$ |
| TiN | No reaction | No reaction |
| TaN[b] | No reaction | No reaction |
| AlN | No reaction | No reaction |
| SiC | No reaction | $MgB_2$ with altered cell size |

[a] $ZrO_2$ is present in the YSZ before reaction.

[b] $TaN_{0.8}$ is present in the TaN before reaction.

Table V: The crystal structure and lattice constants of $MgB_2$ and several well–known substrates.

| Sub. | Crystal system | a (Å) | c (Å) | Surface (lattice constant: Å) | |
|---|---|---|---|---|---|
| $SrTiO_3$ | Cubic | 3.905 | | (100) | Square (3.905) |
| MgO | Cubic | 4.21 | | (100) | Square (4.21) |
| Si | Cubic | 5.431 | | (100) | Square (5.43) |
| | | | | (111) | Hex. (3.84) |
| Sapphire ($Al_2O_3$) | Hex. | 4.76 | 12.99 | C | Hex. (4.76) |
| | | | | R | Rectangle ($4.76 \times 15.39$) |
| SiC | Hex. | 3.081 | 15.12 | (001) | Hex (3.081) |
| $MgB_2$ | Hex. | 3.086 | 3.522 | (001) | – |

Table VI: Physical properties of various types of superconducting junctions of $MgB_2$.

| Ref. | Authors | Material | Junctions | $\Delta$ (meV) | Others |
|---|---|---|---|---|---|
| 44 | Carapella *et al.* | Ex-situ annealed films | $Nb/Al_2O_3/Al/MgB_2$ | 0.9 and 2 (2 gap) | dc- and ac- Josephson effects at 4.2K |
| 45 | Burnell *et al.* | Ex-situ annealed films | $MgB_2$ FIB | - | Ion damage, Ga implantation |
| 46 | Brinkman *et al.* | In-situ annealed films | $MgB_2$ nanobridge | - | dc-SQUID at 15K |
| 47 | Zhang *et al.* | $MgB_2$ flake | $MgB_2/MgB_2$ point contact | ~2.1 | dc-SQUID at 19K |
| 48 | Kohen *et al.* | Bulk (polycrystal) | $Au/MgB_2$ point contact | 3 - 4 | |
| 49 | Schmidt *et al.* | Bulk (polycrystal) | $Au/MgB_2$ point contact | 4.3 - 4.6 | |
| 8 | Szabo *et al.* | Bulk (polycrystal) | $Cu/MgB_2$ point contact | 2.8 and 7 (2 gap) | |
| 50 | Gonnelli *et al.* | Bulk (polycrystal) | $Au(Pt)/MgB_2$ point contact | 2.6 | |
| 51 | | | $MgB_2$ break junctions | - | dc- and ac- Josephson effects |

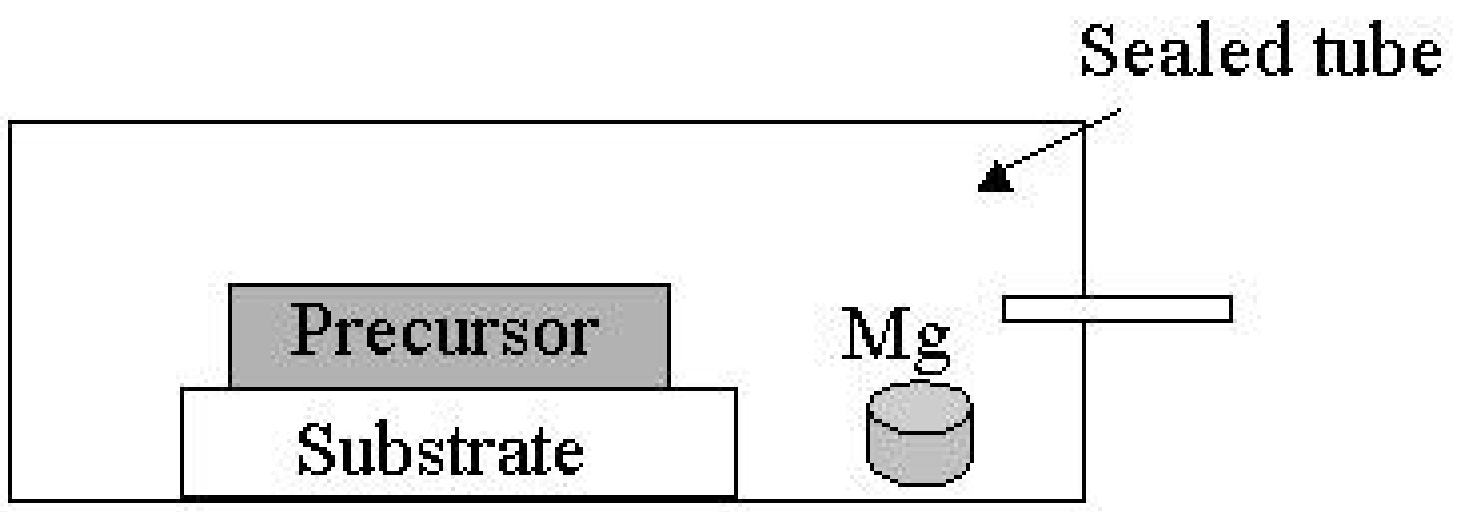

**Fig. 1**

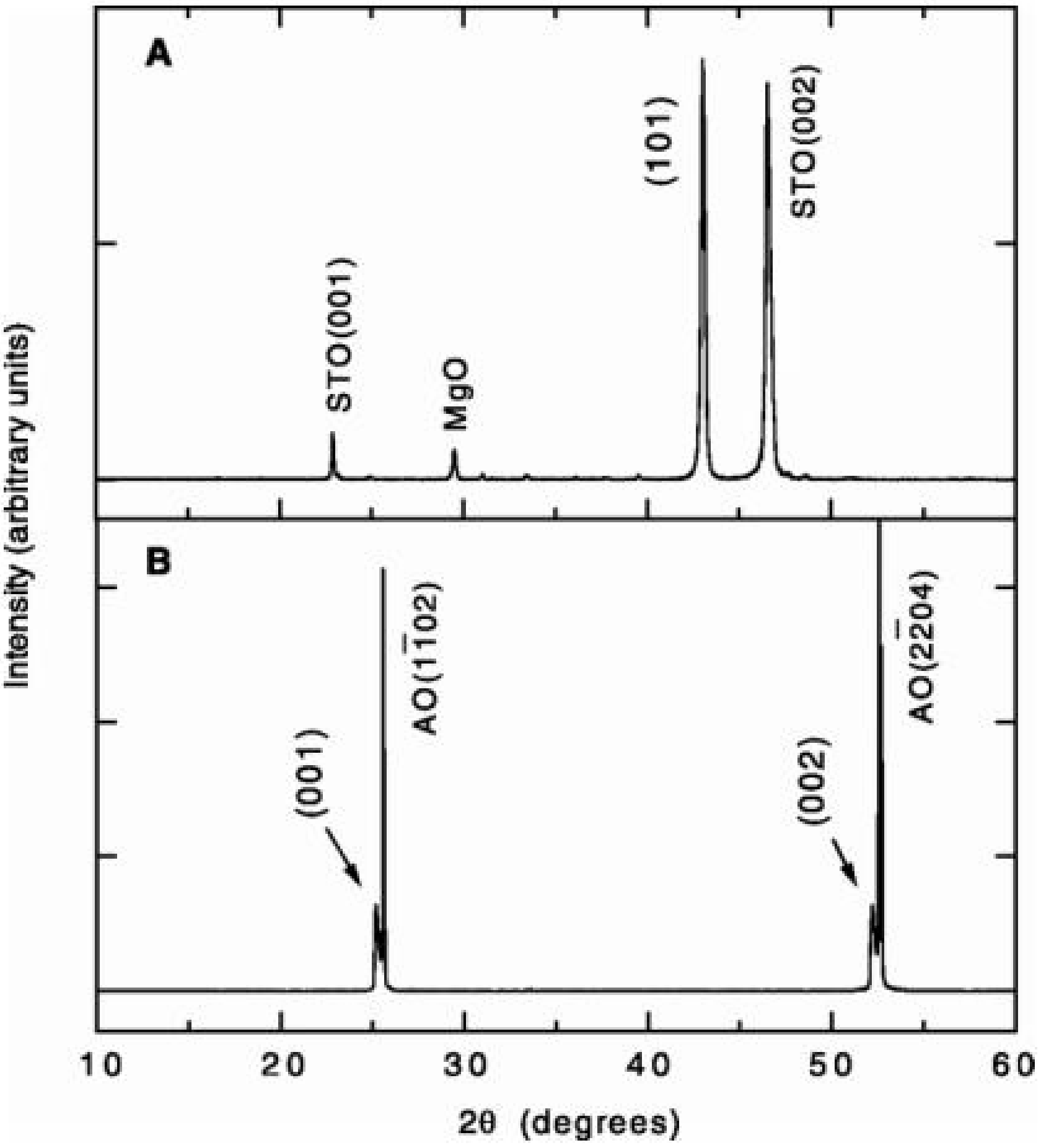

**Fig. 2**

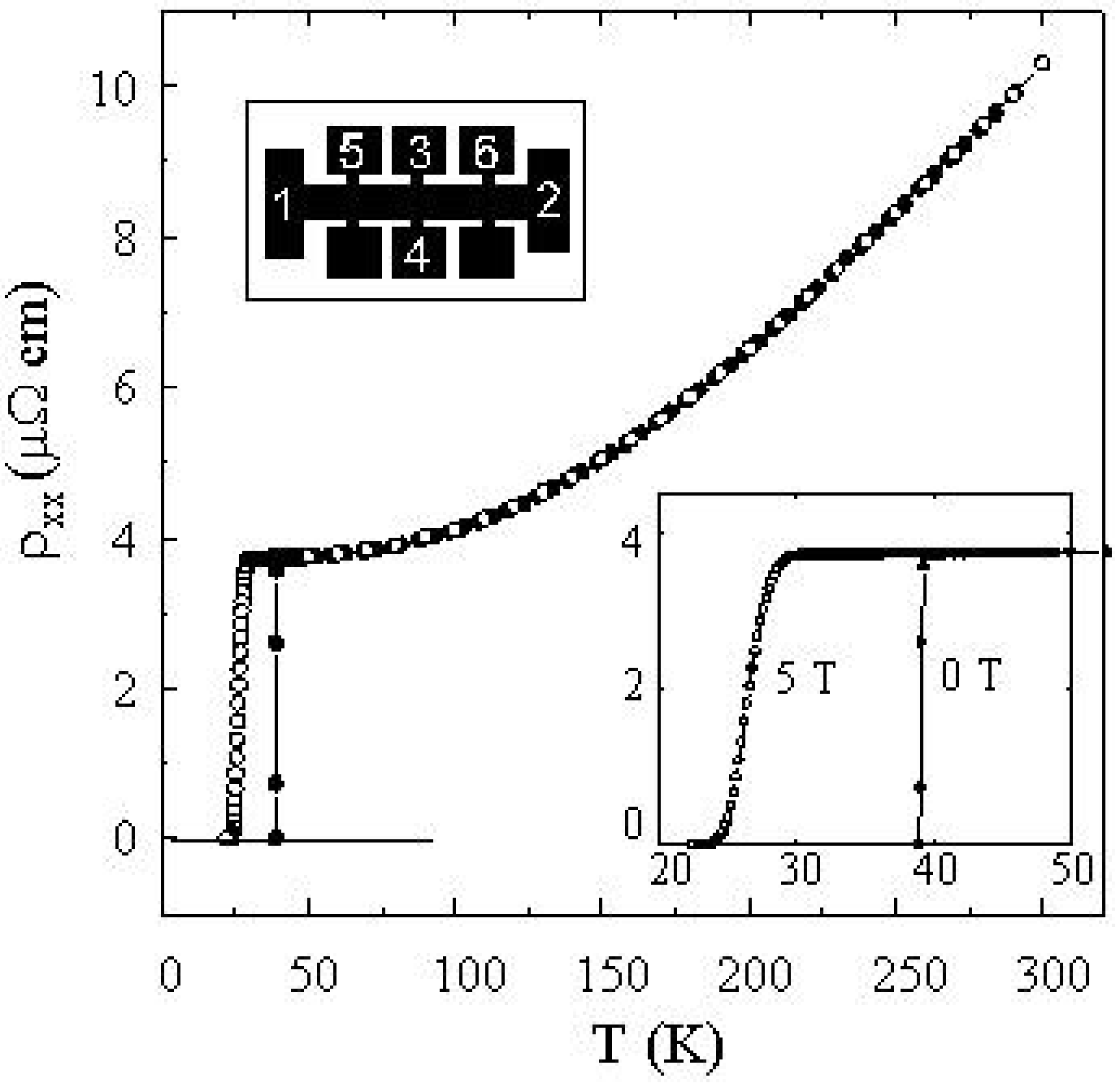

**Fig. 3**

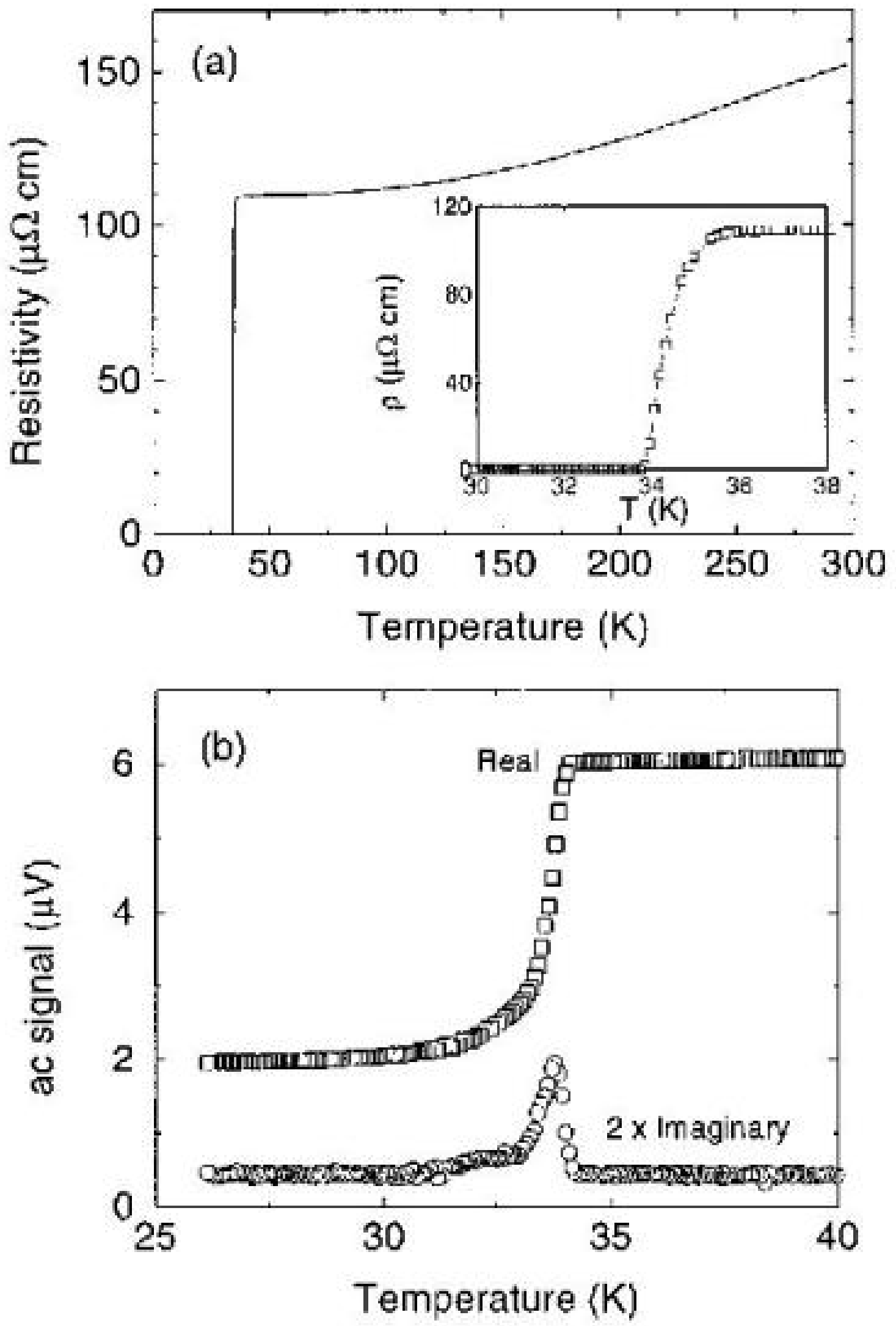

**Fig. 4**

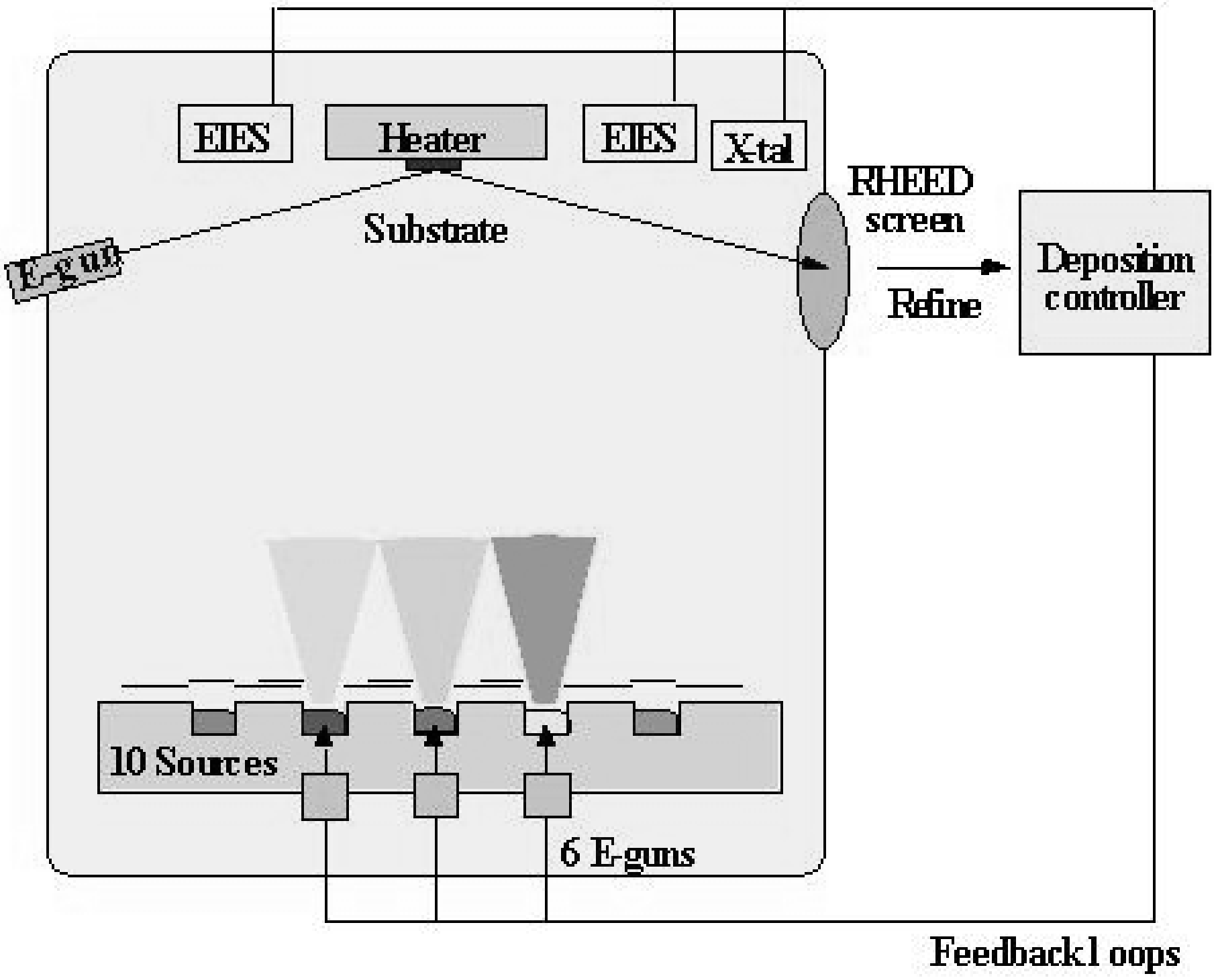

**Fig. 5**

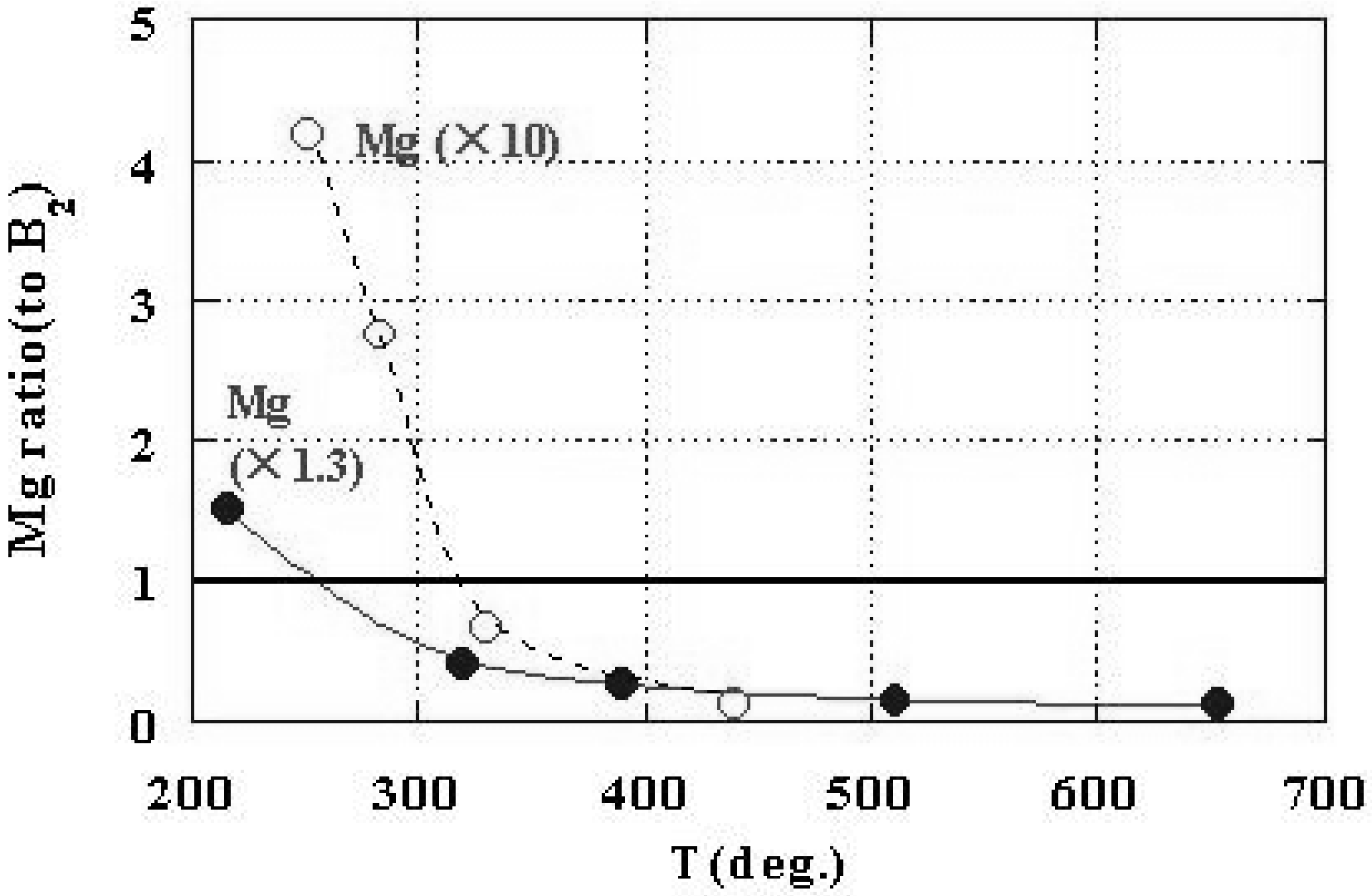

**Fig. 6**

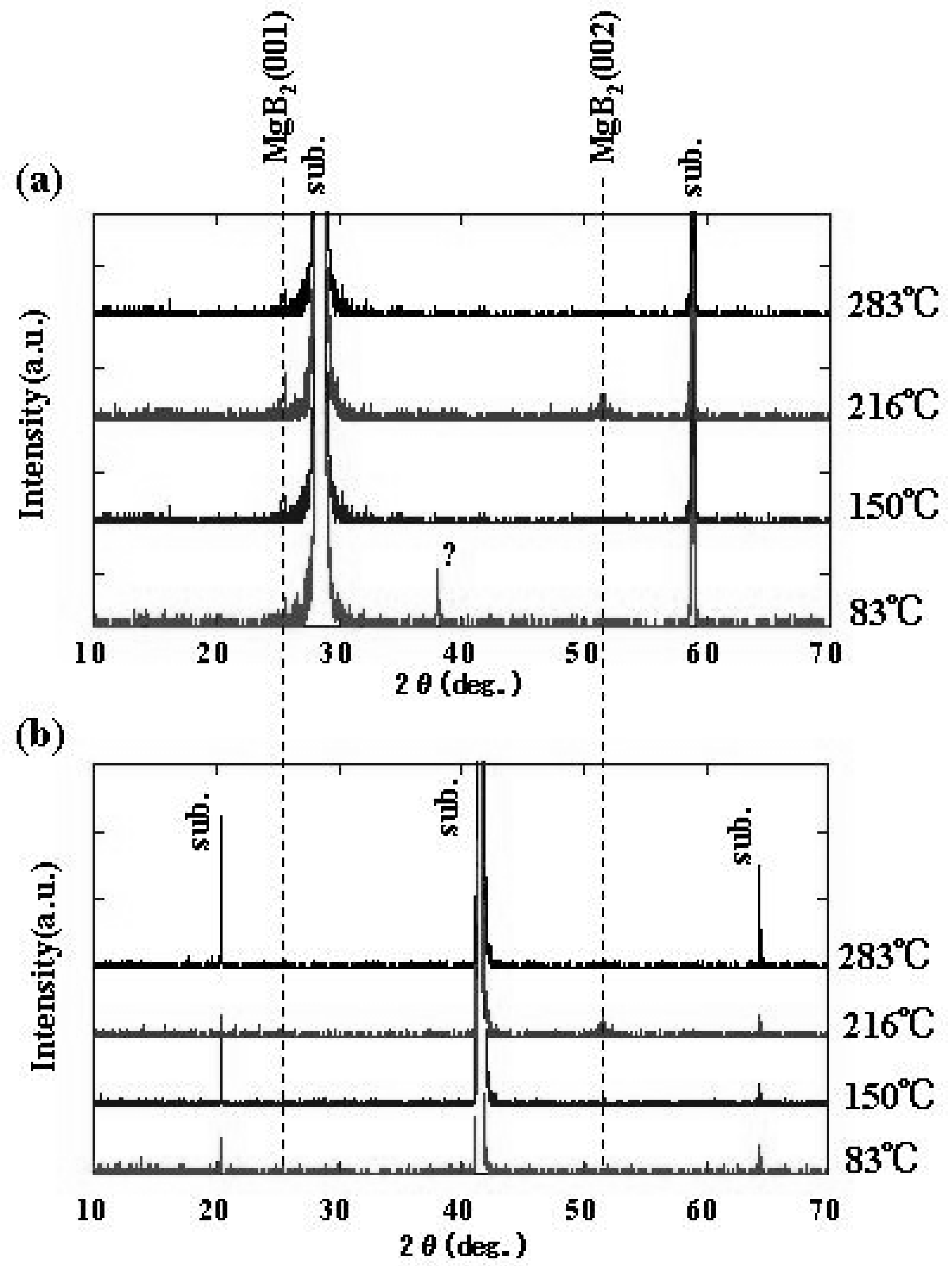

**Fig. 7**

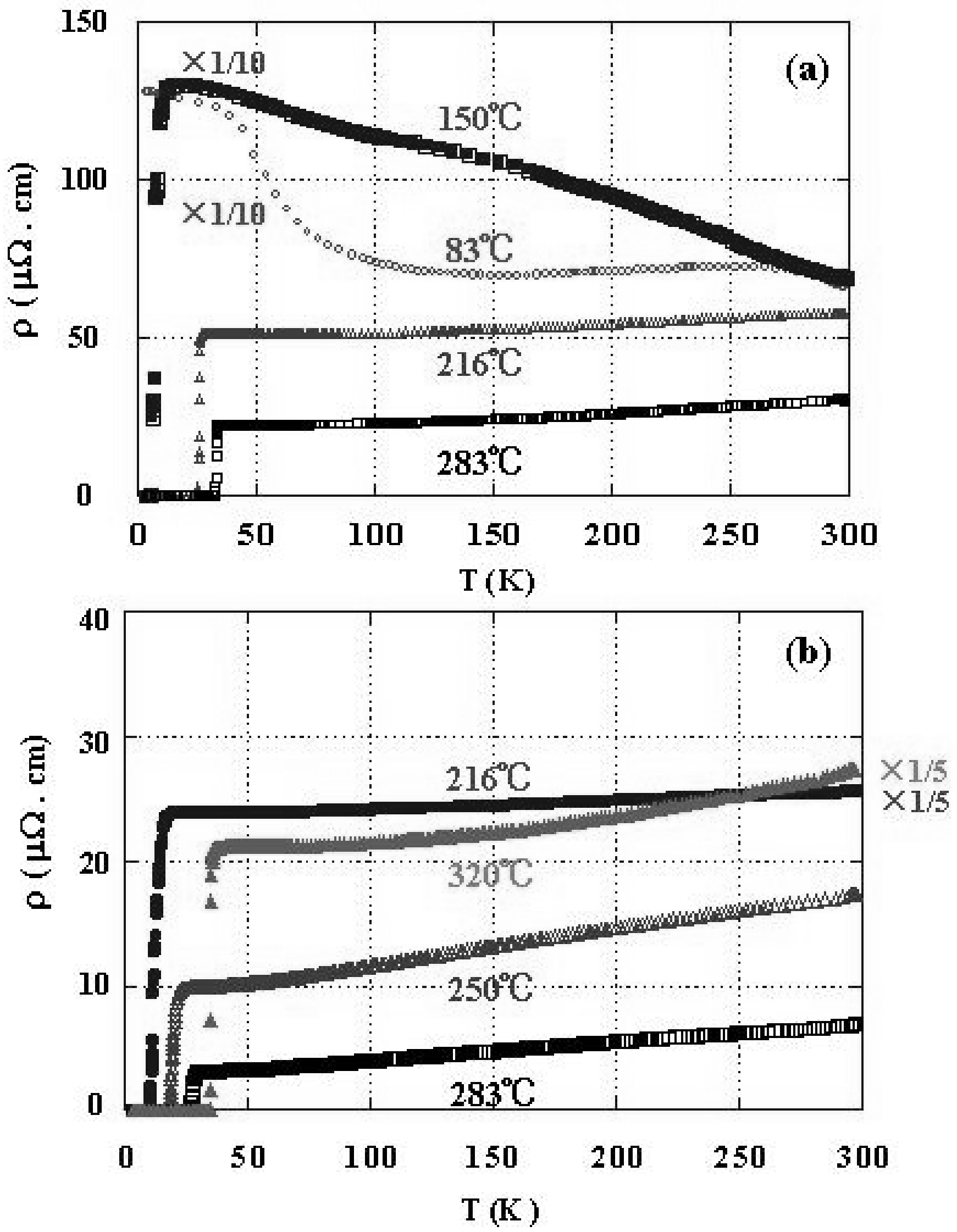

Fig. 8

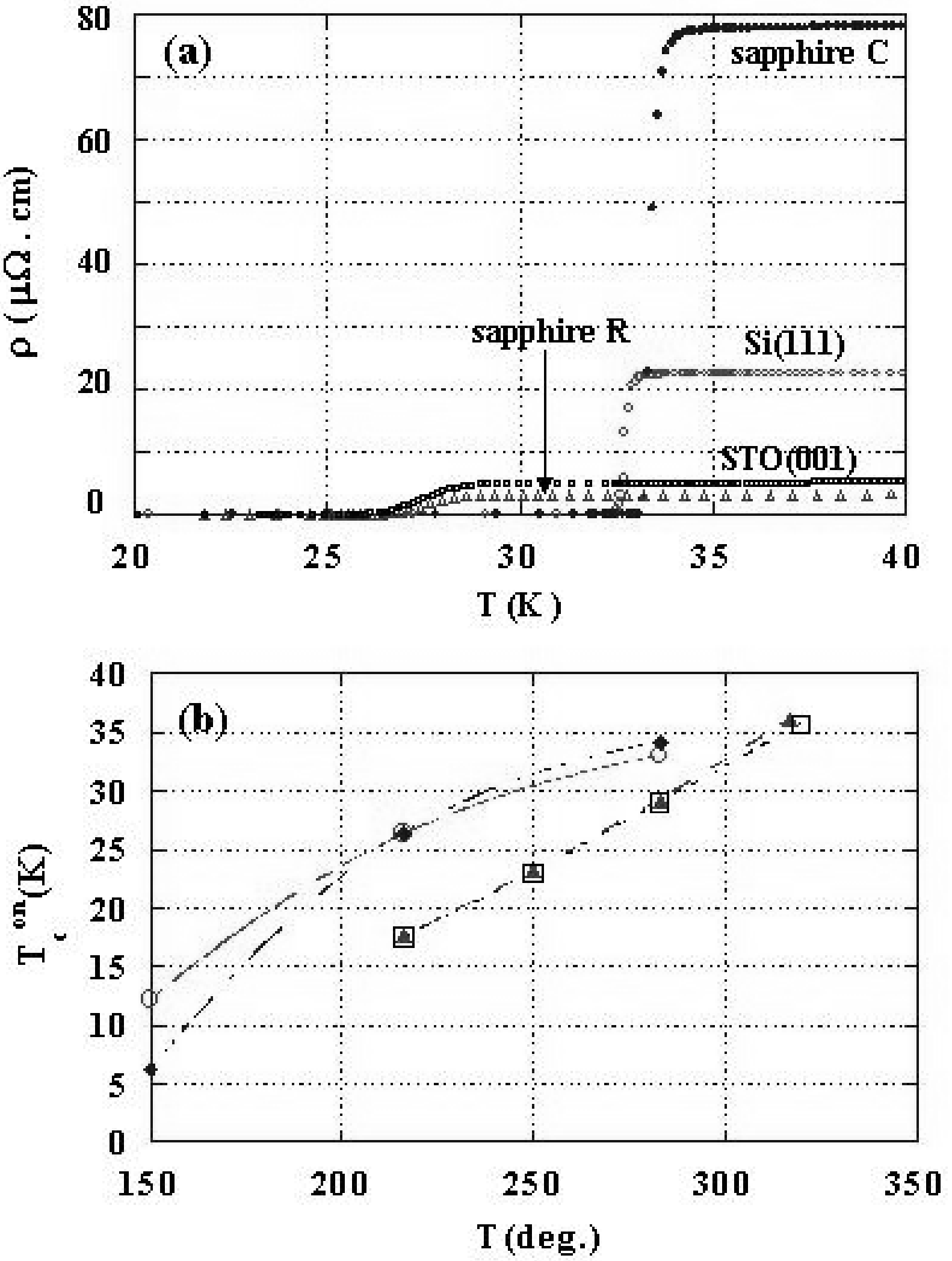

Fig. 9

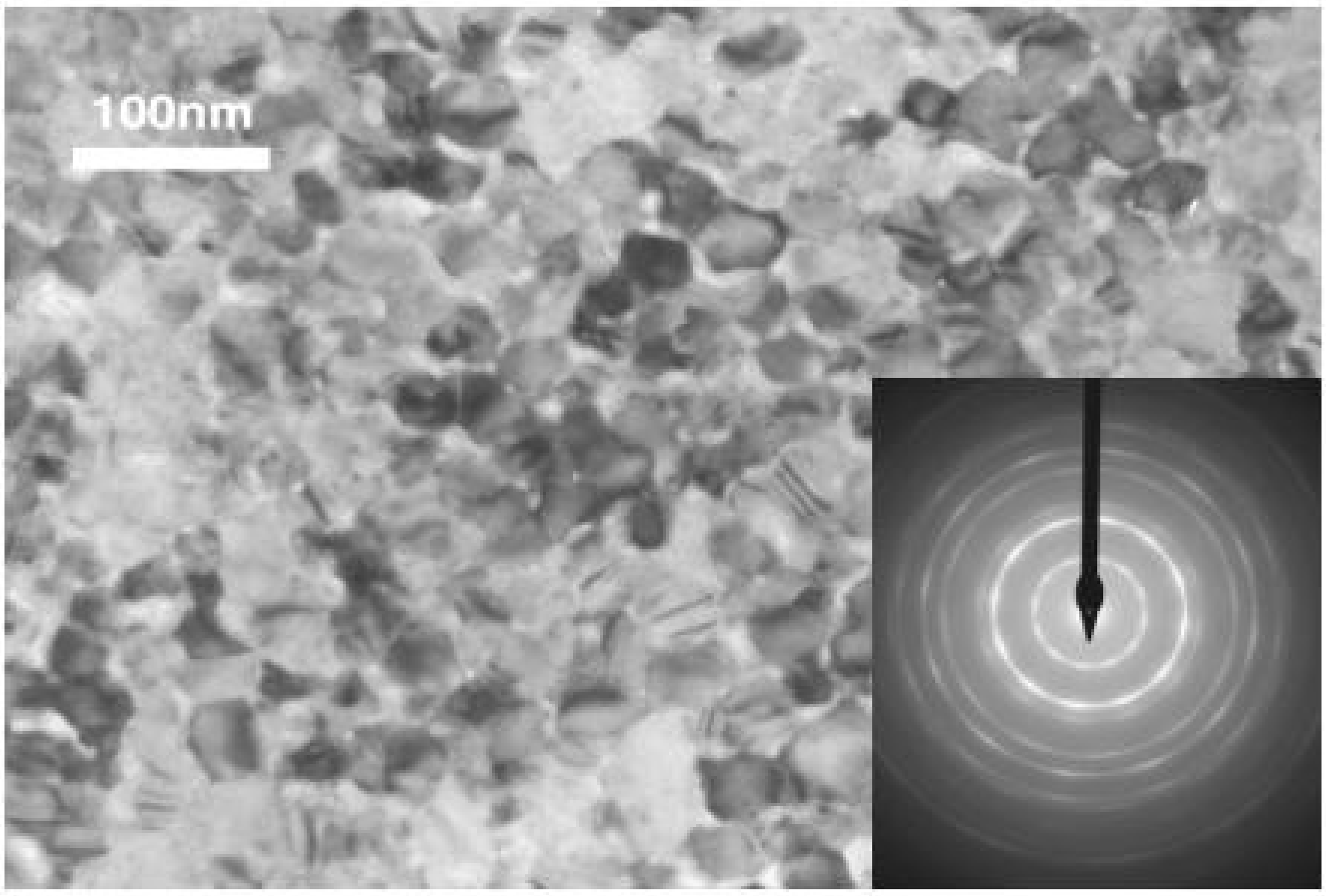

**Fig. 10**

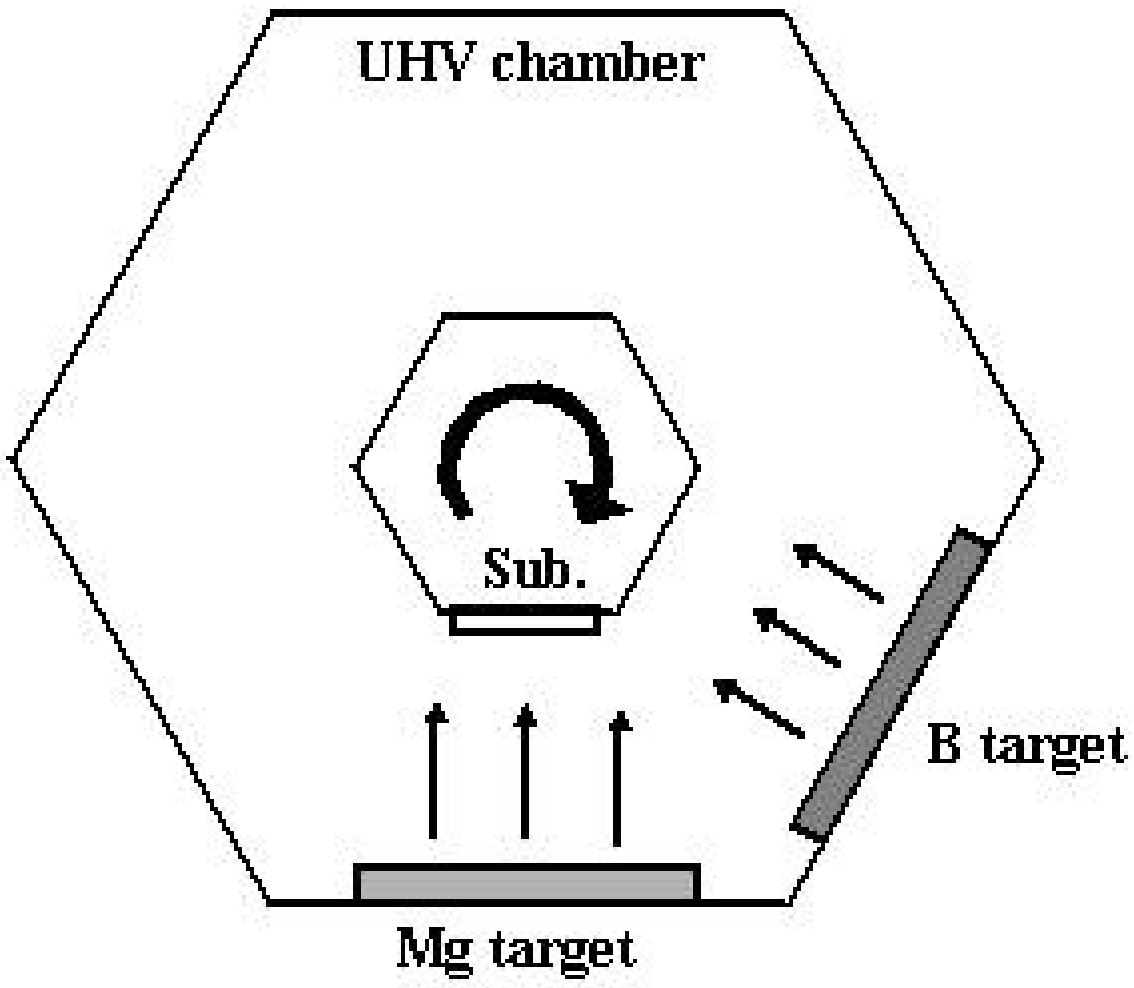

**Fig. 11**

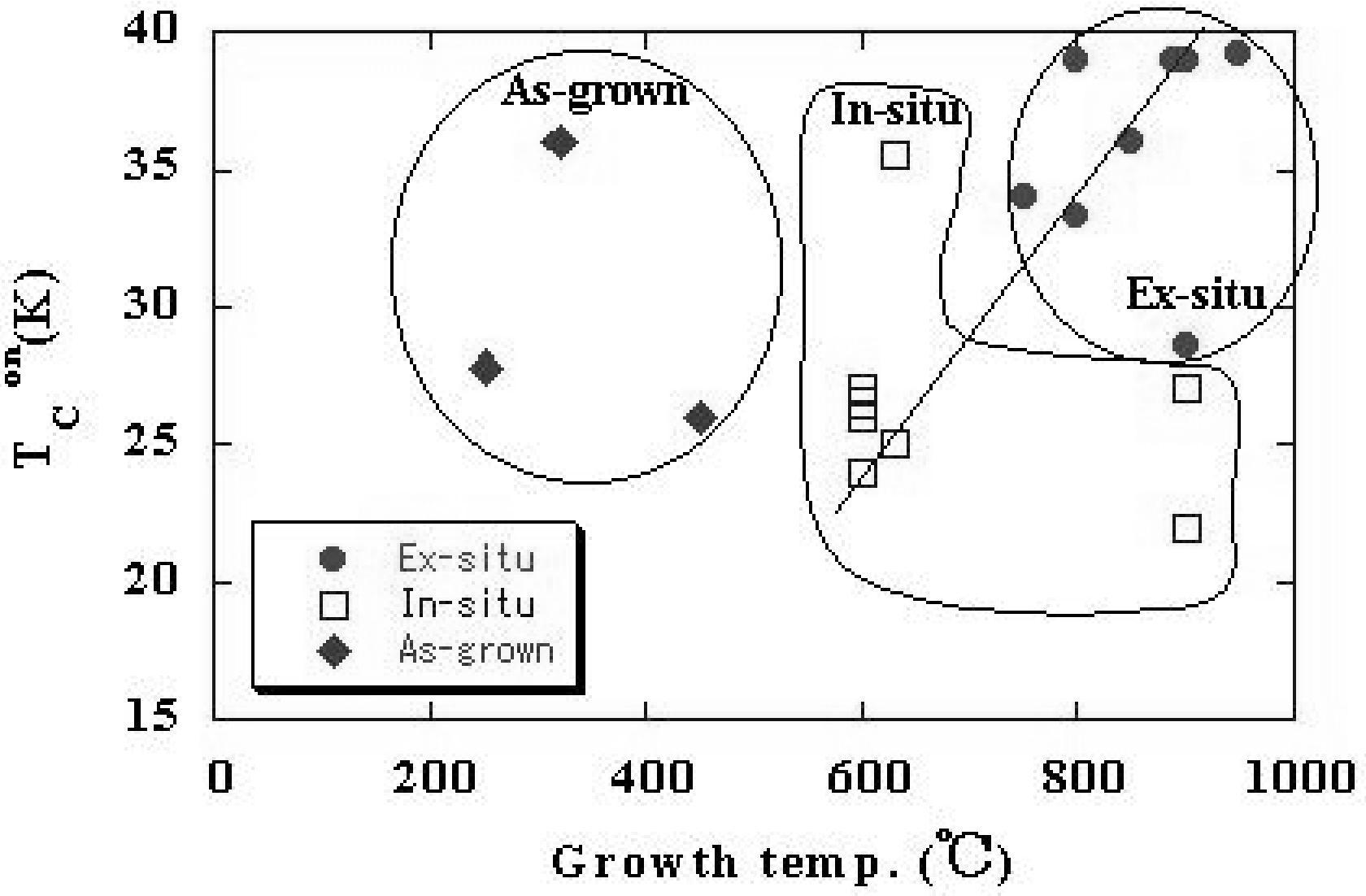

**Fig. 12**

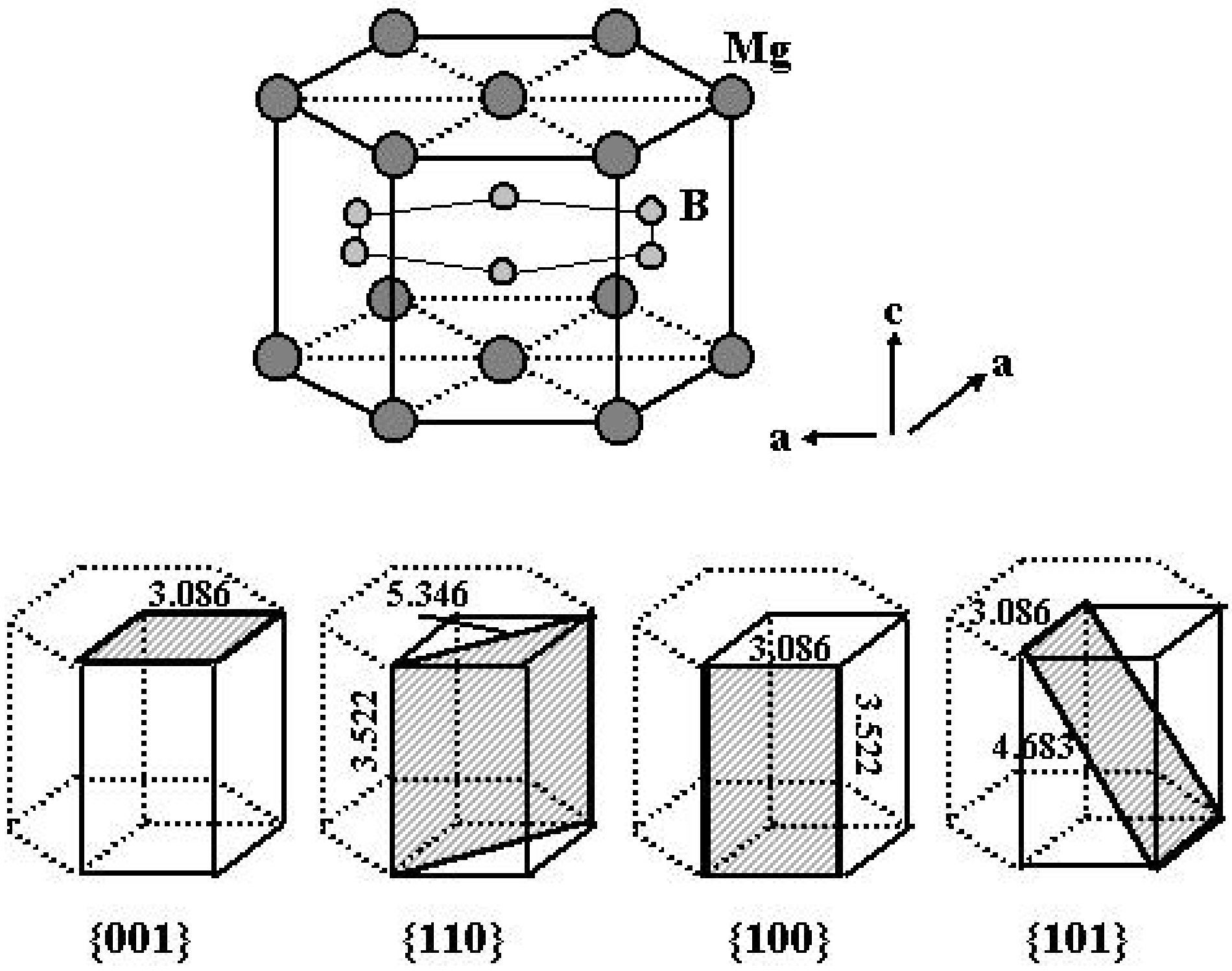

**Fig. 13**

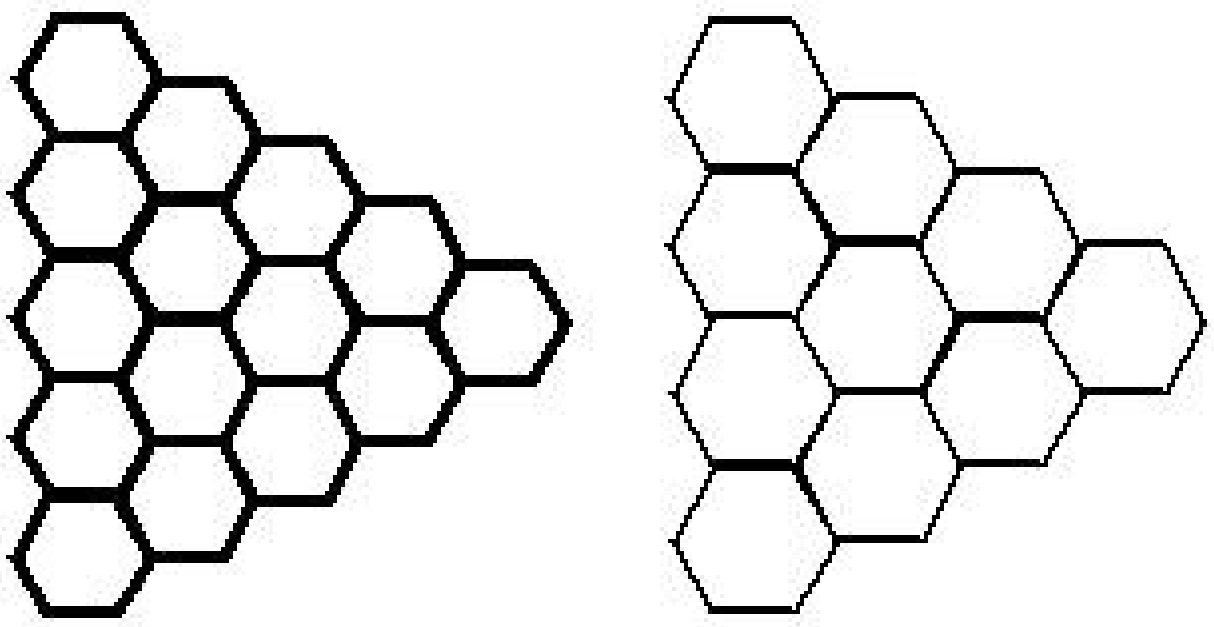

**$MgB_2$ (001): 5 × 5** **Si (111):4 × 4**

**Fig. 14**

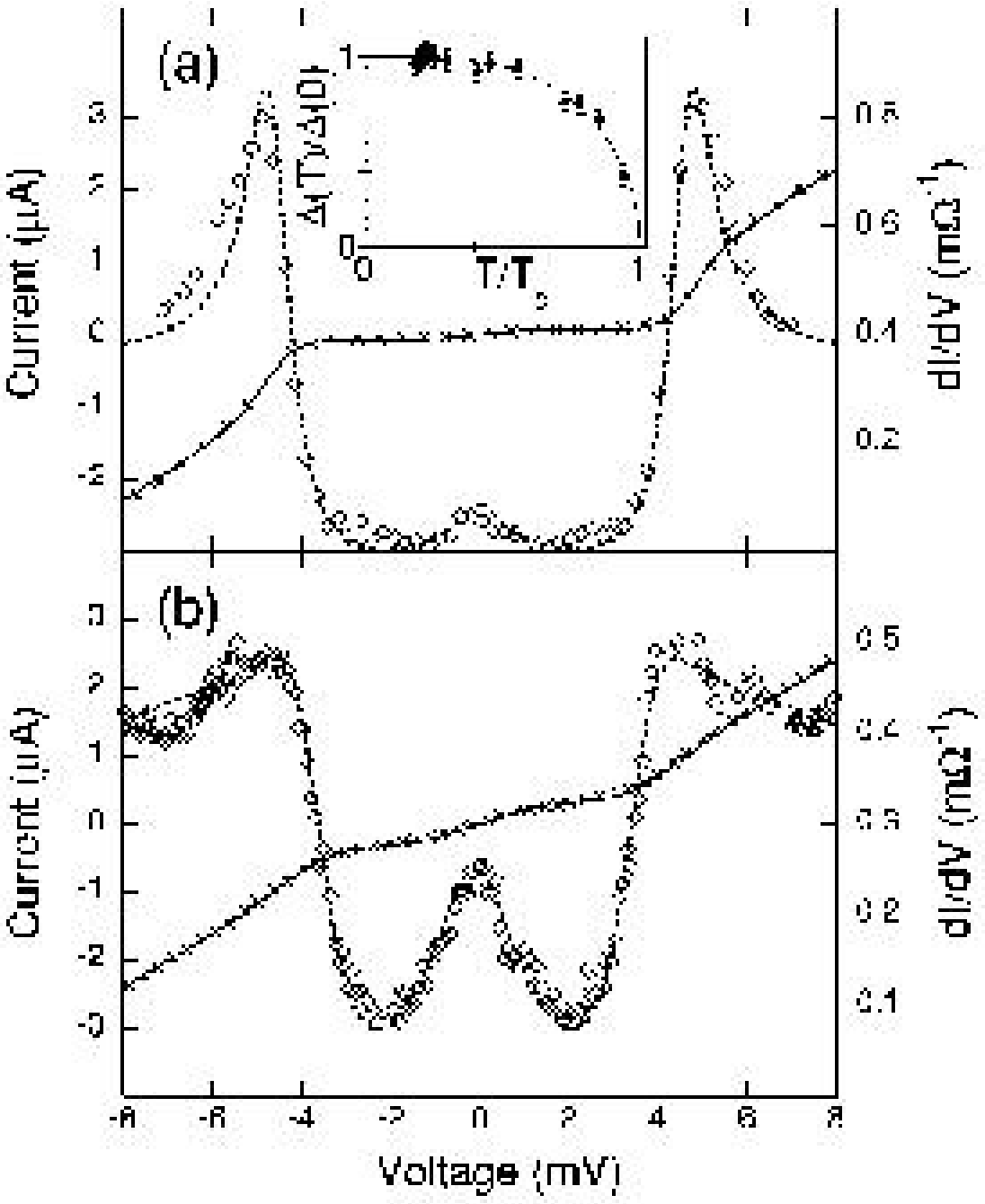

**Fig. 15**

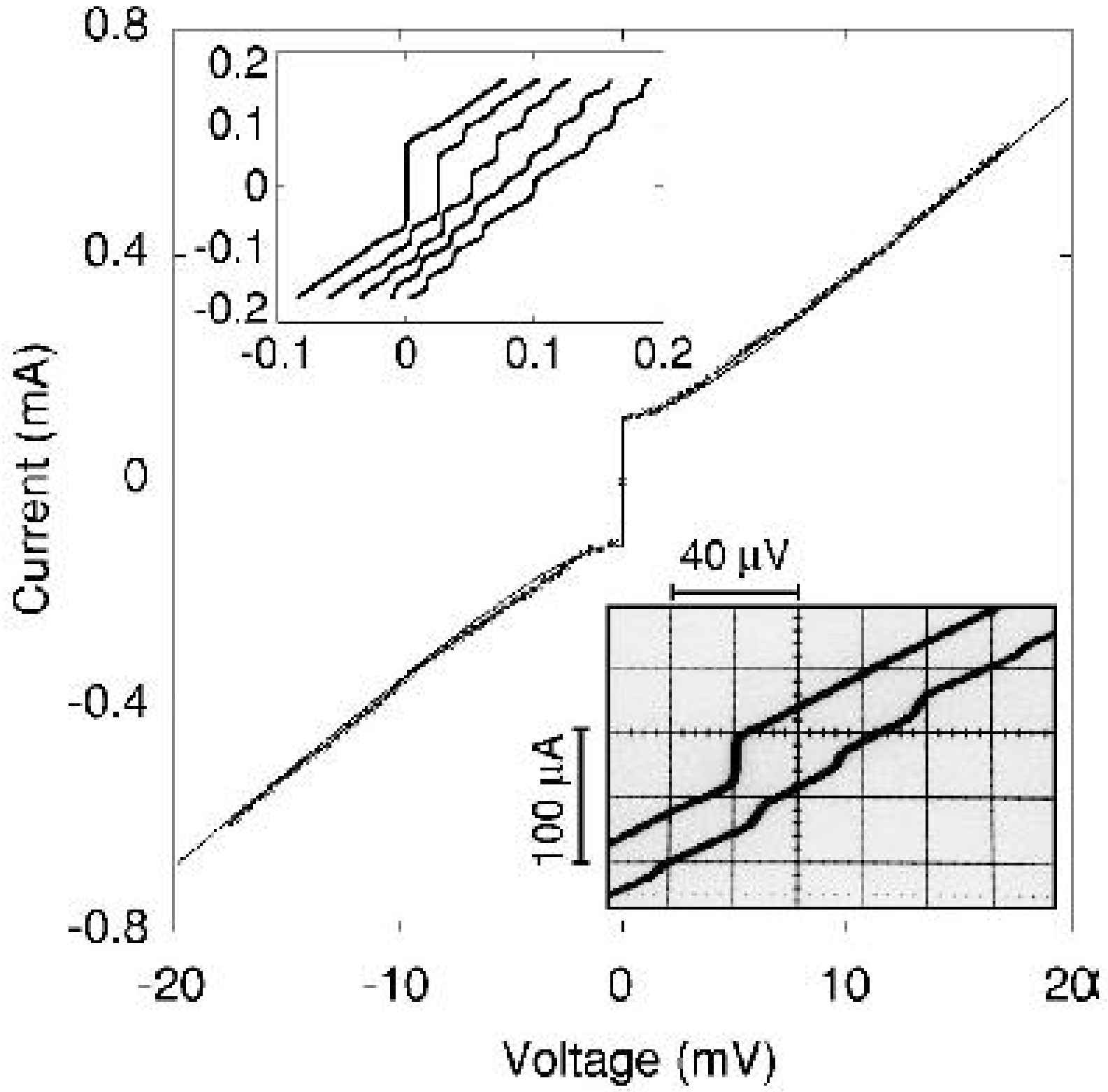

**Fig. 16**